# Liquid crystal self-assembly of upconversion nanorods enriched by depletion forces for mesostructured material preparation

Yong Xie,[a,b] Yuanyuan Li,[b] Guoqing Wei,[a] Qingkun Liu,[a] Haridas Mundoor,[a] Ziyu Chen[b] and Ivan I. Smalyukh[*a,c]

[a] Department of Physics, Material Science and Engineering Program, Department of Electrical, Computer, & Energy Engineering, and Liquid Crystal Materials Research Center, University of Colorado, Boulder, Colorado 80309, USA.
[b] Department of Physics, Beihang University, Beijing 100191, China.
[c] Renewable and Sustainable Energy Institute, National Renewable Energy Laboratory and University of Colorado, Boulder, Colorado 80309, USA.
*Ivan.Smalyukh@colorado.edu

**Monodisperse rod-like colloidal particles are known for spontaneously forming both nematic and smectic liquid crystal phases, but their self-assembly was typically exploited from the fundamental soft condensed matter physics perspective. Here we demonstrate that depletion interactions, driven by non-adsorbing polymers like dextran and surfactants, can be used to enrich self-organization of photon-upconversion nanorods into orientationally ordered nematic and smectic-like membrane colloidal superstructures. We study thermodynamic phase diagrams and demonstrate polarization-dependent photon upconversion exhibited by the ensuing composites, which arises from the superposition of unique properties of the solid nanostructures and the long-range ordering enabled by liquid crystalline self-organization. Finally, we discuss how our method of utilizing self-assembly due to the steric and electrostatic interactions, along with attractive depletion forces, can enable technological uses of lyotropic colloidal liquid crystals and mesostructured composite materials enabled by them, even when they are formed by anisotropic nanoparticles with relatively small aspect ratios.**

## I. Introduction

Colloidal suspensions of anisotropic particles have been the subject of continued research and many exciting discoveries for over 100 years, owing to their rich variety of liquid crystalline phases and fascinating phase transitions.[1] Most of the studies on orientational ordering, optical properties and phase behaviors of colloidal liquid crystals (LCs) were performed on suspensions of rods,[2–7] disks[8,9] or flexible chains[10–12]. The physical behavior of these soft matter systems is commonly understood on the basis of the Onsager's model,[13] which predicts a phase transition from isotropic disordered to nematic ordered phase as the concentration of anisotropic colloidal particles exceeds a critical concentration value. This transition emerges as a result of the entropy maximization and the competition between the entropic contributions associated with the translational and orientational degrees of freedom: as the orientational ordering emerges, parallel arrangements of anisotropic colloidal objects lead to a decrease in orientational entropy and an increase in positional entropy, with an overall maximum of entropy corresponding to a disordered state below critical concentrations and to an orientationally ordered state at concentrations above the critical value. The introduction of attractive depletion interactions,[14,15] driven by non-adsorbing polymers like dextran and micelle-forming surfactants, enriches the phase behavior, as well as often leads to various metastable states and co-existing mesophases. For example, Lekkerkerker and colleagues[16] investigated mixtures of colloidal boehmite rods and flexible polymer, showing that bi- and triphasic phase regions can be obtained. Dogic et al.[17,18] conducted quantitative studies of isotropic-nematic phase transition in a binary suspension of rodlike fd virus (aspect ratio of ~ 130) and dextran polymer, concluding that the order parameter was determined solely by the rod concentration. Savenko et al.[6] studied numerically the phase behavior exhibited by a binary mixture of colloidal hard rods and non-adsorbing polymer, proposing that the nematic and smectic phases were more stable for larger depletants when considering the effective many-body interactions.

During the last decade, many superstructures induced by depletion attractions were also reported for various metallic nanorods[19–21] and semiconductor nanorods[3,22,23]. Most of the previous studies[16–18] focused on understanding of the depletion forces on LC phase behavior of colloidal particles with high aspect ratios, so that, at low depletant concentrations, it could be still understood on the basis of modifications of the Onsager theory, with the order parameter of the nematic phase determined mainly by repulsive electrostatic and steric interactions, while attractive interactions providing structureless cohesive energy. However, how these depletion forces interplay with steric and electrostatic interactions for particles with relatively low aspect ratios and nontrivial geometric shapes (e.g. hexagonal) remains unknown. Moreover, to the best of our knowledge, the influence of depletion forces on the LC mesophase behavior, stability and diversity of self-assembled colloidal superstructures and emergent properties and material behavior of versatile upconversion nanorods (UCNR) remain unexplored, though it could be potentially useful for obtaining a host of mesostructured composite materials with interesting optical properties.

Although the upconversion nanoparticles have attracted considerable attention due to their unique properties of converting near-infrared light to visible or ultraviolet light rather efficiently,[24,25] their use in the design and fabrication of mesostructured composite materials is still limited. This is partially due to the limited means of forming composite materials that contain such nanoparticles organized at the mesoscopic scales and can still maintain the upconversion properties of individual nanoparticles. Doping of solid crystal matrices of UCNRs with rare earth ions, such as Erbium ($Er^{3+}$) and Thulium ($Tm^{3+}$), can mediate the upconversion processes,[26] making them attractive for applications in solar energy harvesting,[27] efficient theranostics,[28] color displays,[29] optical data storage,[30] and so on, but it is often hard to preserve these unique properties when forming bulk composite materials. For example, although luminescence of individual UCNRs can be polarized,[25,31] this property is not preserved in conventional disordered colloidal dispersions and could be maintained only in the dispersions of UCNRs in anisotropic orientationally ordered hosts like thermotropic LCs, although only relatively small percentages of such particles could be dispersed in thermotropic LCs so far.[24] The deployment of the UCNRs and other photon up-converting nanostructures in various applications requires the development of strategies for forming composite materials that inherit properties of the individual nanostructures, in which large concentrations of nanoparticles are also commonly desired.

In this work, we have synthesized monodisperse $Tm^{3+}$ doped UCNRs with different, relatively small aspect ratios, so that the lyotropic LC phase behavior is tuned to occur at rather high volume fractions. We show polarized upconversion luminescence and other emergent properties of the ensuing mesostructured lyotropic LC composites and solid films. We find that the depletion interactions can be used to self-organize UCNRs into orientationally ordered nematic and smectic-like membrane colloidal superstructures, as well as the solid films obtained upon evaporation of the solvent while preserving orientational ordering. The insights into the nanoscale ordering in the ensuing mesostructured composites are provided by the direct scanning electron microscopy (SEM) imaging, revealing a surprising diversity of structures and the evidence for metastable multi-axis orientational ordering. Furthermore, in addition to the colloidal LC composites, our dispersions with relatively low volume fractions of solvent can be readily converted to thin solid films that preserve both the orientational ordering and upconversion properties. Thus, our findings may impinge on the development of mesostructured composites with pre-engineered physical behavior and properties for a host of technological applications.

## II. Experiments

**Chemicals and materials**

All chemicals were used as received without further purification. All organic solvents, dextrans, Yttrium chloride hexahydrate ($YCl_3$ $6H_2O$), Ytterbium chloride hexahydrate

($YbCl_3 \cdot 6H_2O$), Thulium chloride ($TmCl_3$), Gadolinium chloride hexahydrate($GdCl_3 \cdot 6H_2O$), ammonium fluoride ($NH_4F$), sodium fluoride (NaF) and oleic acid (OA) were all obtained from Sigma-Aldrich. Sodium hydroxide (NaOH) was purchased from Fisher Chemicals.

**Synthesis of UCNRs**

Synthesis of β-$NaYF_4$ doped with $Tm^{3+}$, $Yb^{3+}$, $Gd^{3+}$ (UCNR1): In a typical procedure, a deionized (DI) water solution (1.875 mL) of 0.375 g NaOH was mixed with 6.25 mL of ethanol and 6.25 mL of oleic acid under stirring. Then 2.5 mL of rare earth (RE) chlorides $RECl_3 \cdot 6H_2O$ (0.3 M, RE = 33 mol% Y, 40 mol% Gd, 3 mol% Tm, and 24 mol% Yb) and 1.25 mL of $NH_4F$ (2 M) were added under vigorous stirring. After 20 minutes, the solution was transferred into a Teflon-lined autoclave and heated to the elevated temperature of 210 ℃, at which it was kept for 2 h. In the end of this process, the obtained nanorods were collected by centrifugation, washed with ethanol and water several times, and then re-dispersed in cyclohexane. For removing the oleic acid layer capped on the RE ions doped β-$NaYF_4$ nanorod and re-dispersing the UCNRs to an aqueous solution, an acidic solution was prepared first by mixing a small amount of hydrochloric acid (HCl) in DI water and stirred vigorously for 2 h. Following this, the nanoparticle dispersion in cyclohexane was mixed with the mild acidic solution of pH ~ 4. During this process, the OA ligands attached to the surface of the particles are released into the cyclohexane as the free OA molecules and UCNR particles with protonates (charged) surfaces are transferred into DI water. The UCNR particles transferred to the aqueous medium were then collected by centrifugation and dispersed in ethanol or DI water.[31,32]

Synthesis of β-$NaYF_4$ doped with $Tm^{3+}$,$Yb^{3+}$ (UCNR2): In a typical synthesis process, NaOH (0.7 g, 17.5 mmol), OA (90 wt.%, 7.1 g, 22.6 mmol), and ethanol (10.0 g, 21.7 mmol) were thoroughly mixed at room temperature to obtain a white viscous solution. An aqueous solution of NaF (0.58 M, 12.45 mL, 7.20 mmol) was added to it with vigorous stirring until a translucent solution was obtained. Then 80 mol% $YCl_3 \cdot 6H_2O$, 2 mol% $TmCl_3 \cdot 6H_2O$, 18 mol% $YbCl_3 \cdot 6H_2O$ in aqueous solution (1.5 mL, 1.2 mmol) was poured into the above solution under vigorous stirring. After aging for 20 min, the mixture was transferred to a 50 mL teflon-lined autoclave, and heated to the elevated temperature of 230 ℃, at which it was kept for 12 h. The obtained rods were collected by centrifugation, washed with ethanol several times, and finally transferred to the aqueous solution using the same transfer method as that described above.

**Preparation of lyotropic LCs of UCNRs**

UCNR concentration C (mg/μL) was identified and controlled by weighing the particles in a solid powder and then adding the corresponding volume of DI water. The density of hexagonal β-$NaYF_4$ crystal, used in these analyses, is $\rho = 4.2$ g/cm$^3$. In a typical process, a 1.5-milliliter tapered plastic centrifuge tube was weighted first. The prepared UCNR solution was then pipetted into this tube and centrifuged. After that, the supernatant was moved out completely, so that a very viscous, white gel was left in the tube bottom and dried at 40 ℃ for hours. The particle weight can be estimated by subtracting the net weight of the tube from the final total weight of the tube with the dried powder of particles. The corresponding volume of DI water was added into the tube and then dispersed by using ultrasonic bath until a well-dispersed solution was obtained. Different concentrations of dextran were then added to the UCNR colloidal dispersions. For keeping the UCNR concentration a constant, for example, 5 μL known-concentration dextran was pre-prepared and added into the UCNR solution (for example, 10 μL). And then 1 μL water + 4 μL dextran, 2 μL water + 3 μL dextran, 3 μL water + 2 μL dextran, 4 μL water + 1 μL dextran, and 5 μL water were also prepared and added respectively into the UCNR solution with the particles concentration all the same. After sonication, the solution with the same UCNR concentration but different dextran concentrations were prepared and enable for the next use.

**Sample preparation and optical characterization**

Glass cells were typically prepared by sandwiching a 1 mm-thick glass slide and a 0.15 mm-thick coverslip with 15 μm spherical $SiO_2$ spacers (from EPRUI Biotech Co. Ltd.). The UCNR-dextran solution was pipetted into the cell and sealed quickly by a nail polish glue at the cell periphery (the glue was immiscible with the UCNR-dextran aqueous solution). Optical bright-field and POM images of the samples were obtained using a upright optical microscope BX-51 (Olympus) equipped with 10×, 20×, 50× dry objectives (all from Olympus) with numerical aperture NA = 0.3-0.9, a CCD camera (Spot 14.2 Color Mosaic, from Diagnostic Instruments, Inc.), polarizers, and a 530 nm full-wavelength retardation plate. Fluorescence microscopy spectra of UCNRs were obtained using an Olympus IX-71-inverted optical microscope equipped with an appropriate dichroic mirror and spectral filters separating the infrared excitation light from the emission. Nanoparticles were excited using the 980 nm continuous-wave output from a diode laser (from Dragon Lasers) focused into the sample using a 100× oil immersion objective. The upconverted light from the sample was collected by the same objective and sent through suitable optical filters before detected by a color CCD camera (Pointgrey, FLEA-COL) or a spectrometer (USB2000-FLG, Ocean Optics). An analyzer was introduced into the optical path immediately before the spectrometer, thereby controlling the linear polarization of the luminescence signal transmitted to the spectrometer and detector. The confocal microscopy imaging of the samples was performed using an inverted microscope (IX-81 Olympus). In this characterization method, a linearly polarized 980 nm pulsed output from a Ti: Sapphire oscillator (140 fs, 80 MHz, Chameleon Ultra II, Coherent) scans the sample with the help of a set of two Galvano mirrors, defining the lateral imaging area of the sample. The luminescence collected from the particles in an epi-detection mode using a 100× oil immersion objective (NA = 1.4) was sent through a pinhole, confocal with the objective's focal plane, before being detected by a photomultiplier tube (PMT, H5784–20, Hamamatsu).

**SEM and TEM imaging**

For SEM imaging, the colloidal dispersions with nematic and smectic superstructures of self-assembled UCNRs were drop-casted on a clean silicon wafer while letting the solvent evaporate in a programmable climate chamber at a fixed temperature of 18 °C and humidity gradually varied from 45% to 85%. These conditions were optimized for preserving the orientational ordering of the rod-like particles while taking advantage of the fact that the used initial concentrations of nanoparticles were rather high (~50%). During this preparation process, the small gradual reduction of solvent concentration was sufficient to eliminate particle mobility and obtain immobile orientationally ordered assemblies of the nanoparticles, as monitored through polarizing optical observations and as needed for the preparation of orientationally ordered films with polarization-dependent photon upconversion properties. The images of these assemblies, as well as individual nanoparticles, were acquired with a Hitachi S-4800 microscope operating at 10 kV. For TEM imaging, the UCNRs were prepared by casting and naturally drying dilute dispersions of UCNRs in DI water on a formvar-coated copper grid. The TEM images were acquired with an FEI Tecnai G2 F30 microscopy at 300 kV.

## III. Results and discussion

**UCNR characterization**

We have synthesized rod-shaped $\beta$-NaYF$_4$ particles by the modified hydrothermal synthesis methods. Our experiments utilized particles with dimensions 121.2 × 18.6 nm (UCNR1) and 1.6 × 0.17 μm (UCNR2), with the length-width aspect ratios 6.5 and 9.3, respectively (Fig. 1a,d and Fig. S1), deliberately tuned to be below 10. When illuminated with a 980 nm laser light, these particles emit visible light, as shown in Fig. 1d inset, indicating the photon upconversion property. Fig. 1b shows high magnification transmission electron microscopy (TEM) image of the side-profile of UCNR1. A representative high-resolution TEM (HRTEM) image of the particle is shown in Fig. 1c. The clear lattice fringes indicate the formation of the single-crystalline structure with a good crystallinity and the measured lattice spacing of about 0.3 nm corresponds to the

(002) plane determined from the hexagonal phase of β-NaYF$_4$ with a space group of P63/m.[33,34] The doped trivalent lanthanide ions Tm$^{3+}$, Yb$^{3+}$, Gd$^{3+}$ are dispersed as guest atoms in the β-NaYF$_4$ host matrices. SEM image of the edge-on rods is shown in Fig. 1e, based on which the hexagonal cross-section of these rod-like particles is confirmed. Combining with the high-magnification TEM of Fig. 1f, a schematic full-view of the UCNR2 can be presented as in Fig. 1f inset. Fig. 1g shows the HRTEM image of UCNR2. The atom arrays indicate high crystallinity and the lattice spacing of about 0.3 nm of the hexagonal β-NaYF$_4$. The size histograms indicate a rather narrow size distribution of the prepared UCNRs (Fig. S1). The elemental composition of the UCNR2 is measured by energy dispersive X-ray (EDX) microanalysis (Fig. S2). It confirms the presence of F, Na, Y, Tm and Yb, with the molecular formula evaluated to be Na$_{0.66}$Yb$_{0.24}$Tm$_{0.03}$Y$_{0.84}$F$_4$. Since the UCNR2s have length larger than the diffraction limit of optical resolution, they appear as blurred ellipsoids when imaged on the basis of upconversion-derived luminescence (inset of Fig. 1d), as a result of finite optical resolution.

**Lyotropic LCs of UCNRs and polarized luminescence spectra**

Fig. 2 shows the optical micrographs of LC phases formed by UCNRs dispersed along with the non-adsorbing polymer dextran (molecular weight MW=10,000, Stokes radius Rs=2.36 nm). The size of spherical dextran globules is chosen to be much smaller than the dimensions of the UCNR particles (the effective radius of the smaller UCNR1 particles is 9.3 nm) to induce sufficiently strong depletion forces. Fig. 2a-d show the typical micrographs and behavior of the LC phase of the mixtures. Fig. 2a is a bright-field optical microscopy image, indicating that the UCNRs are dispersed uniformly, without aggregations. Fig. 2b shows the image of nematic LC of the mixtures obtained by a polarizing optical microscope (POM) between crossed polarizers. The bright domains in these POM images represent the birefringent areas with the UCNR orientations differing from the orientations of crossed polarizers (oriented along the image edges). Fig.

2c reveals further the spatial variations of the LC director n in the LC phase, as probed under POM with a retardation plate. The spatially uniform dispersion of the UCNRs is further confirmed by the confocal microscopy imaging when excited with a 980 nm laser source as shown in Fig. 2d. For a comparison, we also provide an example of the behavior for unoptimized conditions in the case when we use micellar surfactant cetyltrimethyl-ammonium bromide (CTAB) as a depletant to make the same phase modulation of UCNR1 (Fig. S3). The LC phases in dispersions containing dextran are spatially homogeneous while the dispersions containing CTAB micelles exhibit small micrometer-size alignment domains, separated by various defects (Fig. S3a inset), instead.

We have measured the polarized upconversion luminescence spectra corresponding to unidirectionally aligned regions of nematic phases of UCNR1 and UCNR2 while rotating an analyzer (A) within 0-360° with respect to n placed immediately before the spectrometer (Fig. 2e,f). Our results indicate apparent polarization dependent emission from nematic regions, similar to the emission properties of an individual UCNR particle arising from the symmetry of the transition dipoles.[24] These measurements are consistent with the unidirectional arrangement of UCNR particles in the studied sample regions. The dominant emission peak at 800 nm, corresponding to the transition $^3H_4 - {}^3H_6$ levels of $Tm^{3+}$ ions present 29% and 18% changes, respectively, corresponding to the dextran-induced LC phases of the two different types of UCNRs (Fig. 2e,f). The emission peak at 780 nm present almost 100% polarized changes. The other emission peaks present relatively weak polarized changes as shown in the spectra. This polarization dependence is caused mainly by the expansion of the crystalline lattice of the UCNR and the distortion of the local symmetry.[34]

**Phase diagrams of UCNR-dextran colloidal dispersions**

Previous studies of model colloidal systems by Dogic and co-authors[17,18] demonstrated that the role of depletion forces and attractive interactions is to modify the phase behavior predicted by Onsager theory mainly by expanding the phase co-existence, although these studies were conducted on rods with an aspect ratio of about 130 and larger. To clarify the influence of dextrans on the formation of lyotropic LC phases by the UCNRs with aspect ratios < 10 and hexagonal cross-sections, we carry out investigations of the phase transition as functions of dextran and colloidal rod concentrations. Because of the relatively small aspect ratios, hexagonal cross-sections and surface charging of UCNRs, Onsager's theory cannot be used to predict the exact values of the volume fractions at which the system should transition from isotropic to biphasic and to nematic states, though it can be used for point-of-reference estimates of such concentration values to guide interpretation and analysis of experimental results, as describe below.

The phase diagrams obtained for UCNR1 and UCNR2 are shown in Fig. 3a and Fig. 3b, respectively, for which four distinct regions can be distinguished: isotropic (I), nematic (N), isotropic-nematic (I-N) coexistence and isotropic-smectic (I-S) coexistence. Apparently, at relatively low dextran concentration, the UCNR can be well dispersed. Since the density of UCNRs (4,200 kg/m$^3$)[35] is substantially larger than that of the solvent with dextran (~1,000 kg/m$^3$), the gravitational length $l_g = k_bT/(\pi R^2 \cdot Lg\Delta\rho)$ due to this density difference $\Delta\rho = 3,200$ kg/m$^3$ and the volume $\pi R^2 \cdot L$ determined by the effective nanorod's length L and radius R is 3.9 mm for UCNR1s and 3.5 μm for UCNR2s, where g = 9.8 m/s$^2$ is the gravitational acceleration constant. In our study, the sample thickness was varied from much smaller than $l_g$ to comparable to it (the cell thickness typically was 15 μm and larger for UCNR1s and 1 μm and larger for UCNR2s). By using samples of several different thicknesses below and comparable to $l_g$, we have ensured that our results are unaffected by concentration gradients. As the concentration of colloidal rods increases, the UCNRs form a nematic phase (Fig. 3c-f), due to the steric and electrostatic interactions.[36] This behavior can be explained qualitatively within the framework of the Onsager's hard-rod theory, though the relatively small aspect ratios of UCNRs preclude the use of this theory for quantitative modelling of our system. We have measured the volume fractions corresponding to the isotropic phase to the biphasic coexistence region and then to the nematic phase at zero concentration of dextran. The critical volume concentration for isotropic to biphasic ($\Phi_I$) and biphasic to nematic ($\Phi_N$) transitions can be compared to what these values would be within the Onsager model, where they are estimated as $\Phi_I = 3.3D/L$ and $\Phi_N = 4.5D/L$, yielding $\Phi_I = 50.8\%$ and $\Phi_N = 69.2\%$ for UCNR1 and $\Phi_I = 35.5\%$ and $\Phi_N = 48.4\%$ for UCNR2 (note that these estimates are used only as points-of-reference). Not surprisingly, LC ordering of the studied UCNRs in the dispersions in pure solvent appears only at rather high concentrations (Fig. 3). However, the I-N and I-S coexistence can appear at relatively low rods concentration with increasing concentration of dextran in our system, which is due to the introduction of the depletion attractions. For example, the I-N coexistence can be realized with volume fraction less than 30% for UCNR1 with aspect ratio 6.5 and less than 23% for UCNR2 with aspect ratio 9.3 by the addition of high concentration dextrans (the initial concentration 232 mg/mL for dextran of MW=10,000). The induced depletion prompt the existence of metastable LC states.

POM micrographs shown in the Fig. 3c-h provide additional insights into the phase behavior of the UCNR1-dextran dispersions. A large-area, uniform nematic region is presented first in Fig. 3c-e while a spatially varying director pattern is shown in Fig. 3f. In addition to the nematic-like structures, many self-assembled metastable smectic-like membranes with enhanced ordering are also found (Fig. 3g,h), which correspond to the regions of diagrams separated by dashed lines. When observed using polarizing optical microscopy with and without the additional phase retardation plate, these membranes appear weakly birefringent within the surrounding isotropic background of the two-phase coexistence region of the phase diagram. The presence of smectic membranes is consistent with previous studies[16-18] of similar phase diagrams of colloidal rods and is further supported by our SEM imaging of sample regions similar to the ones shown in Fig. 3g,h, as we discuss below. These membranes possess clear boundary, which separates them from the bulk isotropic surrounding. Inside some of the membranes, unexpectedly, the rods can also exhibit in-plane orientational ordering with spatially varying directionality of orientation (Fig. 3g) or a uniform directionality (Fig. 3h). The in-plane ordering within some of the membranes (Fig. 3g) varies continuously, without the formation of grain boundaries, as common for LC phases. Insights into the nature of this in-plane ordering will be provided by SEM imaging described below.

To further control the interactions in the system, we add excessive free ions ($Na^+$ and $Cl^-$) to further screen the electrostatic repulsions between the particles.[36,37] The phase diagrams obtained in this case are shown in the same Fig. 3a, b, with the phase boundaries depicted using the red lines. The purely isotropic and nematic parts of the diagram are further suppressed and, accordingly, the I-N and I-S coexistence regions are enlarged due to the reduction in repulsive interactions. Interestingly, because of the relatively high volume fractions of UCNRs in the LC dispersions, the colloidal dispersions can be easily converted to solid films with ordered nanorods while preserving the initial LC ordering (see the SEM imaging data below), which may be of interest for many technological uses of such UCNRs in the form of self-assembled composite materials.

**SEM characterization of the orientational ordering**

More insights into the nematic and smectic membrane ordering of the UCNRs are provided by the direct SEM imaging (Fig. 4), which we describe here for the case of UCNR2 particles as an example. This SEM imaging directly reveals the orientational ordering within the mesostructured composites obtained by drying the colloidal dispersions. Additionally, since we monitor the evolution of polarizing microscopy textures while drying the highly concentrated colloidal dispersions, this imaging also reveals nanoscale organization of nanorods corresponding to different optical textures of colloidal dispersions. Fig. 4a shows the unidirectional nematic-like alignment of the UCNR particles, with the director n along the average orientation of the long axes of rods.

Using the SEM images, the orientational order parameter of UCNRs within the mesostructured composites is estimated by measuring the angles θ between the orientation of long axes of individual rods and their average orientation n in the image (we use >120 rods for each image). When UCNR orientations are probed at the surface of a confining substrate (after removing it), we find that the rod-like particles tend to be oriented parallel to this interface, though deviating from their average orientation n and thus yielding the finite values of order parameter. The two-dimensional order parameter values $S_{2D} = 2<\cos2θ> - 1$ in this case, on the basis of probing it in four different images, are within S = (0.89−0.94). When the SEM images are obtained in the locations corresponding to the bulk of original dispersions, the rod orientations tend to have the uniaxial symmetry, and we estimate $S_{3D} = 1/2(3<\cos2θ> - 1)$ to be within $S_{3D}$ = 0.91-0.95, which indicates the rather high degree of orientational ordering in our system. The high values of order parameter within the mesostructured thin films imaged by SEM are of interest for their practical uses in applications utilizing polarization dependence of upconversion luminescence. Although the drying process can potentially alter the values of the scalar order parameter, it is interesting to note that the high values of the scalar order parameter are also common for colloidal nematics formed by rod-like particles.[14-16]

With adding dextran and then further increasing its concentration, the local transition from nematic to smectic membrane ordering can be also observed, as shown in Fig. 4b and its inset, where the smectic membrane and (in some cases) nematic-smectic coexistence regions can be distinguished. These SEM observations are consistent with optical imaging (Fig. 3c-h). The nematic-smectic coexistance is reminiscent to the so-called "cybotactic cluster", the smectic-like domains observed in the nematic phase of thermotropic LCs, typically in the region close to the transition from nematic to smectic phase.[1,9,38–40] Fig. 4c inset shows a magnified side view of a smectic domain of the UCNRs with in-plane ordering of minority of the rod-like nanoparticles. It reveals that, in addition to the majority of nanorods aligned along the local layer normal $n_a$, some rods align along a direction in the plane of rods $n_b$. The schematics of nematic and smectic ordering provided below the corresponding SEM images (Fig. 4e) qualitatively illustrate these unusual features of the observed colloidal assembly, which is preserved upon drying films. Besides, a transition from isotropic, isotropic-nematic intermediate-state to nematic-smectic coexistence can be also observed when we tune the dextran concentration, as shown in detail in the supplementary Fig. S4, demonstrating the sequence of structural changes within the diagram. Depending on the location within the phase diagram, the fraction of UCNRs aligned orthogonally to the membrane can vary and these in-plane rods can exhibit orientational ordering along $n_b$ (Fig. 4b,c,e). The in-plane ordering of UCNRs in some of the smectic membranes is consistent with the POM micrographs, which show weak birefringence even when the membranes are oriented with their layer normal along the optical axis of the microscope (Fig. 3g,h). Thermal fluctuations of the birefringent regions within such domains, which we observed under optical microscope, indicate that such assemblies are not just aggregates or sediments, but rather LC "fluid" metastable assemblies characteristic. Fig. 4c presents smectic domains with higher ordering, where local arrangements of the rods are often hexagonal, but where defects (vacancies or in-plane rods) appear making it short-ranged.

To our surprise, we also find a two-axis nematic ordering of the UCNR in some regions of the phase diagrams. In this case, in addition to the majority of rods orienting their preferred long axis $n_a$, some also orient along another axis $n_b$ (Fig. 4d,e). The SEM images that further illustrate the diversity of this behavior are shown in the supplementary Fig. S5. The importance of the relative dimensions of colloidal rods and depletant can be appreciated by comparing these structures to the ones formed by the UCNR1 rods (Fig. S6) with relatively larger polydispersity and smaller aspect ratio.

**Discussion**

For colloidal rods with high aspect ratios and circular cross-sections, which were studied previously,[17,18,41] the role of depletants and attractive depletion forces is to provide an effective confinement and expanding the co-existence of the multi-structure and multi-phase regions. Our study of relatively low aspect-ratio UCNRs with hexagonal cross-sections reveals a generally similar but more complex behavior. The overall interactions between UCNRs in the used solvents with depletants are subject to steric interactions, van der Walls (important only at short distances), depletion attractions and electrostatic repulsions, all of which are highly anisotropic. However, for simplicity, the short-range van der Walls forces can be neglected and the relatively short-range electrostatic repulsions (the Debye screening length in our system is estimated to be 10 nm, given that the ionic strength of the solution is < 1 mM) can be partially accounted for by "re-sizing" the effective dimensions of the hexagonal nanorods and then considering their electrostatics-modified steric interactions. The excluded volume of colloidal rods with hexagonal cross-sections depends not only on the orientations of long axes of the rods, but also on the relative orientations of their hexagons (Fig. 5a). Therefore, the geometry of our particle can influence both steric-electrostatic repulsive and depletion attractive interactions. For example, the cumulative excluded volume for two hexagonal UCNRs is minimized when their faces are matched as shown in Fig. 5b. This observation indicates that depletion forces should drive mutual alignment of UCNRs with the equilibrium configurations containing parallel side faces in a hexagonal ordered state. However, one can also imagine how geometric shape of hexagonal and other faceted rods can give origins to metastable colloidal assemblies. For example, two rods with each of the two mutually orthogonal pairs of UCNRs shown in their initial configuration in Fig. 5c cannot reduce the excluded volume through rotations (because of sliding on their edges) until almost perfectly aligned. Therefore, a situation can arise where the combination of electrostatic-steric and depletion forces is strong enough to drive ordering of some of the rods but not the others, thus resulting in metastable colloidal superstructures with multi-axis nematic and smectic membrane orientational ordering, which we observe in experiments (Fig. 3 and 4). Formation of these various colloidal superstructures of the UCNRs, demonstrated by the combination of optical and SEM characterization, may also arise in other colloidal systems of faceted nanorods and other colloidal particles with anisotropic geometrically nontrivial shapes. Furthermore, the emergence of these different forms of orientational ordering is of interest for technological applications as it may allow for preserving the polarization-dependent optical photon-upconverting properties of UCNRs in the bulk of mesostructured materials inheriting these properties.

One of the common strategies of fabricating mesostructured crystalline structures of nanoparticles for technologically desirable materials, such as photonic crystals and various crystalline lattices of plasmonic nanoparticles, involves colloidal self-assembly followed by drying of solvents.[42–45] In these approaches, both positional and orientational ordering are usually retained upon solvent drying, though the drying process often affects the crystal lattice constants or orientational order parameters exhibited by nanoparticles such as gold nanorods. Our findings, presented in this work, show that the orientationally ordered states of LC colloidal assemblies (where no long-range positional order is present before or after solvent evaporation) can be also retained upon preparation of solid mesostructured films through evaporating solvent of lyotropic colloidal LC dispersions while starting at high initial concentrations of particles. Although the order parameter and other characteristics can be affected by the evaporation process, even when optimized to minimally affect the orientational ordering, we have shown that mesostructured composites obtained with this approach can exhibit rather high values of the orientational order parameter >0.9, which makes them of interest for applications whenever there is a need of forming mesostructured composites that inherit anisotropic, polarization-dependent optical properties of nanoparticles. Our findings also indicate that the highest scalar order parameters of nanorods in the solid dried films are achieved when using the initial colloidal dispersions starting from the nematic phase, though partially ordered nematic-like and smectic-membrane-like configurations of nanorods can be also obtained when preparing the initial colloidal dispersions of nanorods in the coexistence parts of the diagram, where dextran and ions can be used for a limited control of orientational ordering while working with varying concentrations of UCNRs.

## IV. Conclusions

We have demonstrated lyotropic LC self-assembly of UCNRs, which stems from repulsive steric and electrostatic colloidal interactions enriched by the geometric shapes of particles and depletion forces due to non-adsorbed polymer dextran. These self-assembled colloidal superstructures exhibit various nematic, bi-phasic and smectic ordering themes, including stable and long-lived metastable configurations with uniaxial ordering or with multi-axis ordering of the rods with hexagonal cross-sections. Our findings not only promote the progress of fundamental nano-materials science and soft condensed matter physics, but also open new possibilities for the design of composite mesostructured materials needed for applications ranging from solar energy harvesting to color displays and optical data storage.

## Acknowledgements

We acknowledge discussions with P. Davidson, S. Park, B. Senyuk, T. Lee and Y. Yuan. This study was partially supported by the U.S. Department of Energy, Office of Basic Energy Sciences, Division of Materials Sciences and Engineering, under Award ER46921, contract DE-SC0010305 with the University of Colorado Boulder (I.I.S., Q.L., G. W. and H.M.) and also by the National Natural Science Foundation of China awards 51502011, 11474015 and 61227902 (Y.X., Y.L. and Z.C.). I.I.S. gratefully acknowledges the hospitality and support of the CNRS (France) during his stay at the University of Paris-Sud in Orsay, France.

## References

1   P. d. Gennes, J. Prost, The physics of liquid crystals 2nd ed. Oxford University Press,

USA: 1995.

2   Q. Liu, Y. Cui, D. Gardner, X. Li, S. L. He and I. I. Smalyukh, Nano Lett., 2010, 10, 1347-1353.

3   L. S. Li, J. Walda, L. Manna and A. P. Alivisatos, Nano Lett., 2002, 2, 557-560.

4   N. R. Jana, L. A. Gearheart, S. O. Obare, C. J. Johnson, K. J. Edler, S. Mann and C. J. Murphy, J. Mater. Chem., 2002, 12, 2909-2912.

5   L. S. Li and A. P. Alivisatos, Adv. Mater. 2003, 15, 408-411.

6   S. V. Savenko and M. Dijkstra, J. Chem. Phys., 2006, 124, 234902-8.

7   J. Kim, A. Cotte, R. Deloncle, S. Archambeau, C. Biver, J. Cano, K. Lahlil, J. Boilot, E. Grelet and T. Gacoin, Adv. Funct. Mater., 2012, 22, 4949-4956.

8   B. Dan, N. Behabtu, A. Martinez, J. S. Evans, D. V. Kosynkin, J. M. Tour, M. Pasquali and I. I. Smalyukh, Soft Matter, 2011, 7, 11154-11159.

9   H. K. Bisoyi and S. Kumar, Chem. Soc. Rev., 2011, 40, 306-319.

10  A. Choudhary, G. Singh and A. M. Biradar, Nanoscale, 2014, 6, 7743-7756.

11  S. Ahn, P. Deshmukh, M. Gopinadhan, C. O. Osuji and R. M. Kasi, ACS Nano, 2011, 5, 3085-3095.

12  B. O. Rivera, T. Westen and T. J. H. Vlugt, Mol. Phys., 2016, 114, 895-908.

13  L. Onsager, Ann. N.Y. Acad. Sci., 1949, 51, 627-659.

14  S. Asakura and F. Oosawa, J. Polym. Sci., 1958, 33, 183-192.

15  A. Takeaki and T. Hajime, J. Phys.-Condens. Mat., 2008, 20, 072101.

16  J. Buitenhuis, L. N. Donselaar, P. A. Buining, A. Stroobants and H. N. Lekkerkerker, J. Colloid. Interf. Sci., 1995, 175, 46-56.

17  Z. Dogic, K. R. Purdy, E. Grelet, M. Adams and S. Fraden, Phys. Rev. E, 2004, 69, 051702.

18  Z. Dogic, Phys. Rev. Lett., 2003, 91, 165701.

19  K. Park, H. Koerner and R. A. Vaia, Nano Lett., 2010, 10, 1433-1439.

20  Y. Xie, Y. J. Liang, D. X. Chen, X. C. Wu, L. R. Dai and Q. Liu, Nanoscale, 2014, 6,

3064-3068.

21  Y. Xie, Y. F. Jia, Y. J. Liang, S. M. Guo, Y. L. Ji, X. C. Wu, Z. Y. Chen and Q. Liu, Chem. Commun., 2012, 48, 2128-2130.

22  D. Baranov, A. Fiore, M. van Huis, C. Giannini, A. Falqui, U. Lafont, H. Zandbergen, M. Zanella, R. Cingolani and L. Manna, Nano Lett., 2010, 10, 743-749.

23  D. V. Talapin, E. V. Shevchenko, C. B. Murray, A. Kornowski, S. Forster and H. Weller, J. Am. Chem. Soc., 2004, 126, 12984-12988.

24  H. Mundoor and I. I. Smalyukh, Small, 2015, 11, 5572-5580.

25  Y. Zhang, L. Zhang, R. Deng, J. Tian, Y. Zong, D. Jin and X. Liu, J. Am. Chem. Soc., 2014, 136, 4893-4896.

26  F. Zhang, J. Li, J. Shan, L. Xu and D. Zhao, Chem. Euro. J., 2009, 15, 11010-11019.

27  C. Li, X. Yang, J. C. Yu, T. Ming and J. Wang, Chem. Commun., 2011, 47, 3511-3513.

28  J.-N. Liu, W.-B. Bu and J.-L. Shi, Accounts Chem. Res., 2015, 48, 1797-1805.

29  R. Deng, F. Qin, R. Chen, W. Huang, M. Hong and X. Liu, Nat. Nano., 2015, 10, 237-242.

30  M. Gu, Q. Zhang and S. Lamon, Nat. Rev. Mater., 2016, 1, 16070.

31  F. Zhang, Y. Wan, T. Yu, F. Q. Zhang, Y. F. Shi, S. H. Xie, Y. G. Li, L. Xu, B. Tu and D. Y. Zhao, Angew. Chem. Int. Ed., 2007, 46, 7976-7979.

32  F. Wang, Y. Han, C. S. Lim, Y. H. Lu, J. Wang, J. Xu, H. Y. Chen, C. Zhang, M. H. Hong and X. G. Liu, Nature, 2010, 463, 1061-1065.

33  A. A. Fedorova, A. I. Fedulin and I. V. Morozov, J. Fluorine. Chem., 2015, 178, 173-177.

34  Q. L. Wu, J. F. Pei and G. J. De, J. Lumin., 2014, 152, 192-194.

35  H. C. Liu, C. T. Xu, G. Dumlupinar, O. B. Jensen, P. E. Andersen and S. A. Engels, Nanoscale, 2013, 5, 10034-10040.

36  H. Mundoor, B. Senyuk and I. I. Smalyukh, Science, 2016, 352, 69-73.

37   J. Polte, Cryst. Eng. Comm., 2015, 17, 6809-6830.

38   C. Zhang, M. Gao, N. Diorio, W. Weissflog, U. Baumeister, S. Sprunt, J. T. Gleeson and A. Jakli, Phys. Rev. Lett., 2012, 109, 107802.

39   V. Borshch, Y. K. Kim, J. Xiang, M. Gao, A. Jakli, V. P. Panov, J. K. Vij, C. T. Imrie, M. G. Tamba, G. H. Mehl and O. D. Lavrentovich, Nat. Commun., 2013, 4, 2635-2643.

40   S. Singh, R. Deb, N. Chakraborty, H. Singh, V. Gupta, K. Sinha, P. Tandon, M. Omelchenko, N. V. Rao and A. R. Ayala, New J. Chem., 2016, 40, 6834-6847.

41   L. Kang, T. Gibaud, Z. Dogic and T. C. Lubensky, Soft Matter, 2016, 12, 386-401.

42   G. von Freymann, V. Kitaev, B.V. Lotsch and G. A. Ozin, Chem. Soc. Rev., 2013, 42, 2528-2554.

43   Y. Liang, Y. Xie, D. Chen, C. Guo, S. Hou, T. Wen, F. Yang, K. Deng, X. Wu, I. I. Smalyukh and Q. Liu, Nat. Commun., 2017, 8, 1410.

44   B. Pietrobon, M. McEachran and V. Kitaev, ACS Nano, 2008, 3, 21-26.

45   X. Ye, L. Jin, H. Caglayan, J. Chen, G. Xing, C. Zheng, V. Doan-Nguyen, Y. Kang, N. Engheta, C.R. Kagan and C. B. Murray, ACS Nano, 2012, 6, 2804-2817.

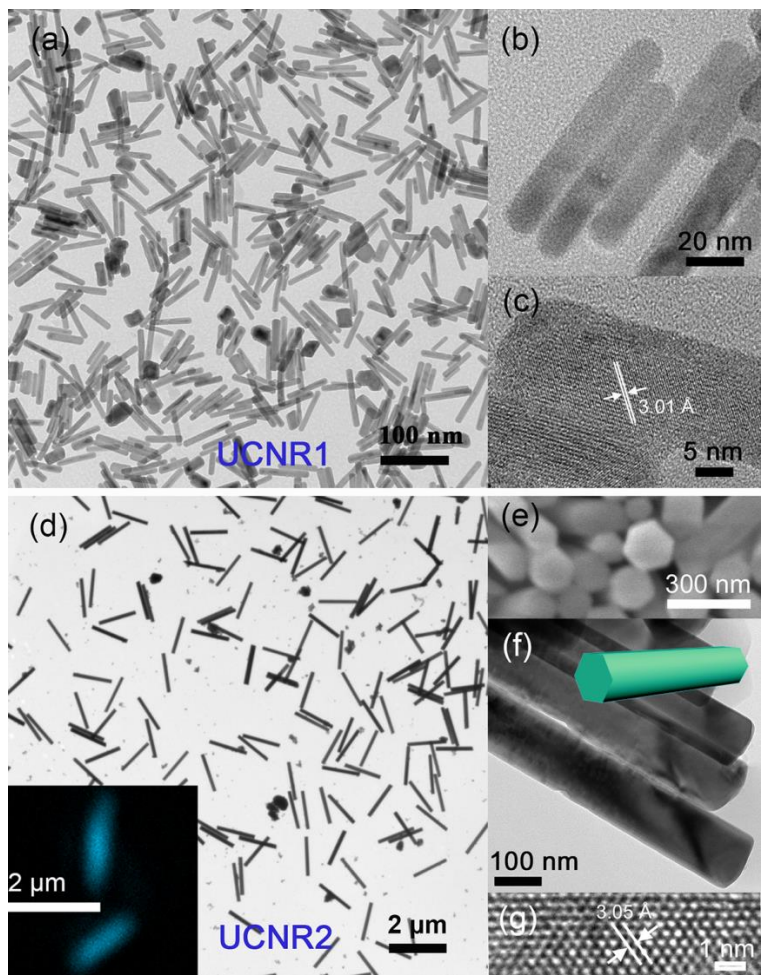

**Fig. 1** Physical characterization of UCNRs. (a) TEM image of the UCNR1 nanoparticles with the average length to diameter aspect ratio of 6.5. (b) High-magnification TEM image and (c) HRTEM image of the UCNR1. (d) TEM image of the UCNR2 with the average aspect ratio of 9.3. The inset shows the upconversion-based fluorescence image of two UCNR2 rods excited by a 980 nm laser, with the fluorescence collected in the epi-detection mode (the color was post-processed) and characterized by a photomultiplier tube. (e) High-magnification edge-on SEM image, (f) side-profile TEM image and (g) HRTEM of the UCNR2. The inset in (f) shows a schematic of the UCNR colored in green to stand-out atop of the image.

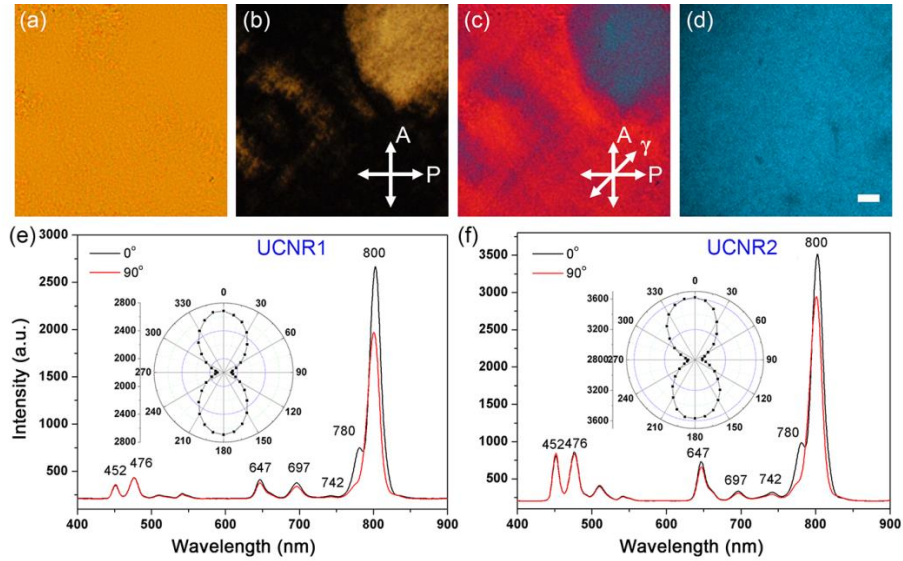

**Fig. 2** Optical images and polarized luminescence spectra of the lyotropic LC phase of UCNRs in solutions with a depletant dextran (MW=10,000, Rs = 2.36 nm). (a) The transmission-mode optical microscopy image. (b,c) POM image taken (b) without and (c) with a 530nm full-wavelength retardation plate with its slow axis marked by γ. (d) The upconversion-based fluorescent microscopy image. The concentrations of the UCNRs and the dextran are about 0.57 mg/μL and 32 mg/mL, respectively. The scale bar is 30 μm. (e, f) The graphs show polarization-dependent emission spectra for the analyzer (A) in the emission detection channel A∥n (0°) and A⊥n (90°) in the uniformly aligned nematic regions of self-assembled colloidal superstructures of UCNR1 and UCNR2, respectively. The insets in (e,f) show polar plots representing the variation of the emission peak intensity at 800 nm corresponding to the rotation of A with respect to n within 0–360°.

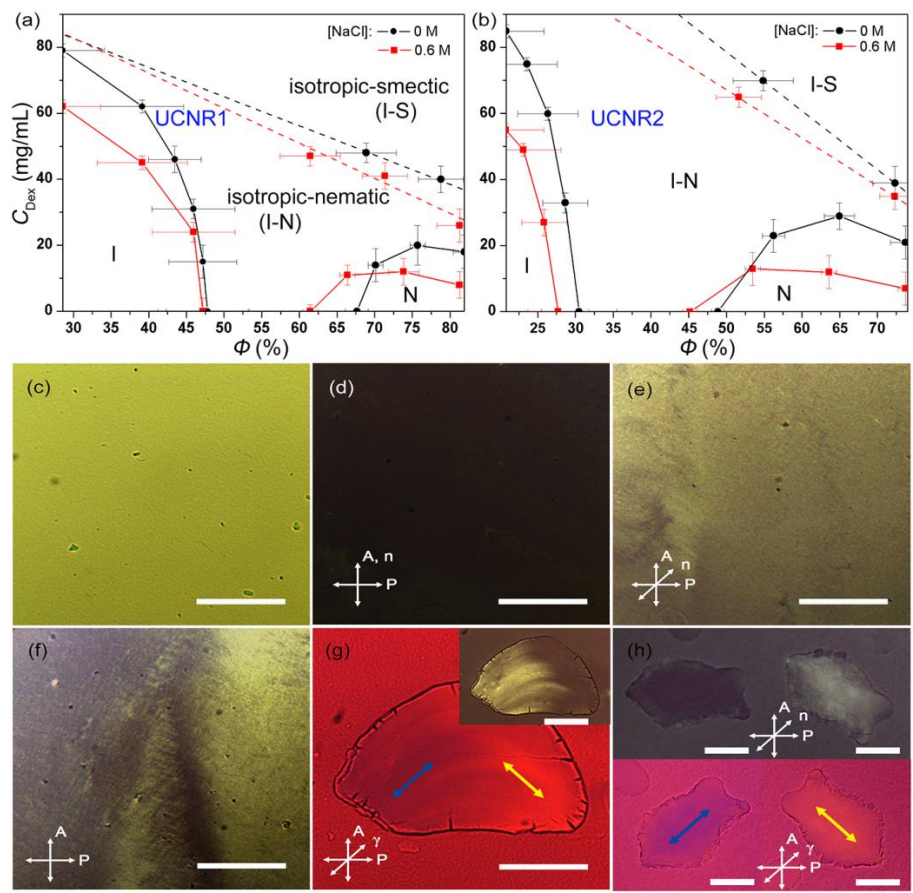

**Fig. 3** Phase diagrams and representative POM images of the UCNR dispersions. (a) Phase diagram for a mixture of UCNR1 and dextran DX1 (MW=10,000, Rs ~ 2.36 nm) with and without NaCl. (b) Phase diagram for a mixture of UCNR2 and dextran DX2 (MW=670,000, Rs ~ 16 nm) with and without NaCl. The red square and black circle symbols are the experimental results with the error bars (determined based on at least 20 independent experiments) shown. The dashed lines indicate the partitions of the diagram between isotropic-nematic (I-N) and isotropic-smectic membrane (I-S) coexistence regions. (c-e) Micrographs of the nematic dispersion samples, where (d) and (e) are POM images obtained with a rotatable sample stage for the sample with n at 0° (the darkest in the tunable range) and at 45°, respectively. The concentrations of the UCNR1 and the dextran DX1 are about 0.65 mg/μL and 20 mg/mL, respectively. (f) A POM image showing the areas of UCNR nematic phase with the spatially distorted director structures. (g) A micrograph of the self-assembled smectic colloidal membrane that nucleates from the isotropic phase in the polymer rich background. The image was taken under POM with the γ plate. The arrows represent the direction of rods in the plane of the membrane, which are a minority of rods as compared to the ones orthogonal to the membrane. The inset is the corresponding POM image obtained without the γ plate. (h) POM images obtained (top) without and (bottom) with the γ plate, revealing the integral ordering of in-plane of the UCNRs membrane. The concentrations of the UCNR1 and the dextran

DX1 here are about 0.6 mg/μL and 46 mg/mL, respectively. The scale bars are 100 μm. The high-resolution and high-magnification SEM images corresponding to these polarizing optical micrographs are shown in Fig. 4 (nematic and smectic membranes) and supplementary Fig. S1c (isotropic structures).

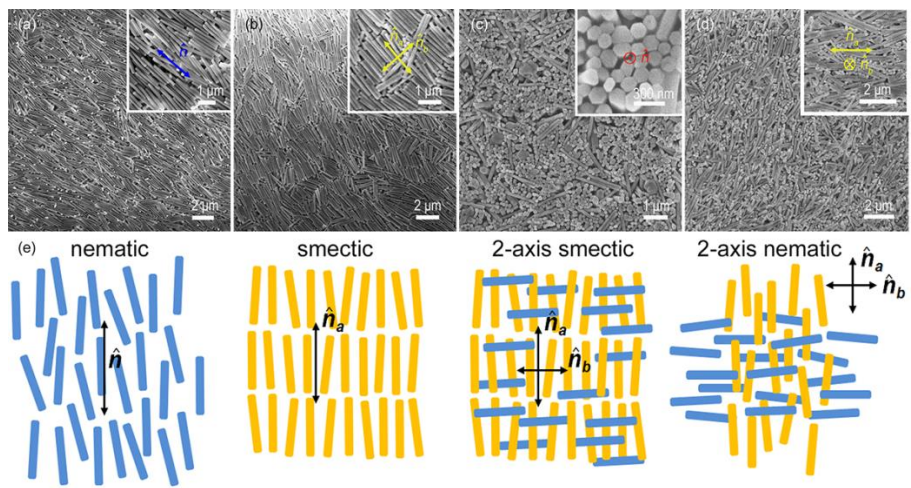

**Fig. 4** SEM images and schematics showing colloidal self-assembled structures. These structures possess (a) nematic ordering, (b) smectic-membrane-like ordering, (c) smectic membranes with two-axis orientational ordering and (d) nematic organization with the two-axis ordering of UCNRs. The used dextran DX2 is characterized by MW=670,000 and Rs=16.9 nm with concentration about 16.7 mg/mL. The initial concentrations of UCNRs are approximately 25, 35, 50 and 65 mg/mL corresponding to the self-assemblies depicted in (a)-(d), respectively. The insets are the enlarged images showing details of the ordering. (e) schematics of the UCNR ordering corresponding to the images shown in (a)-(d).

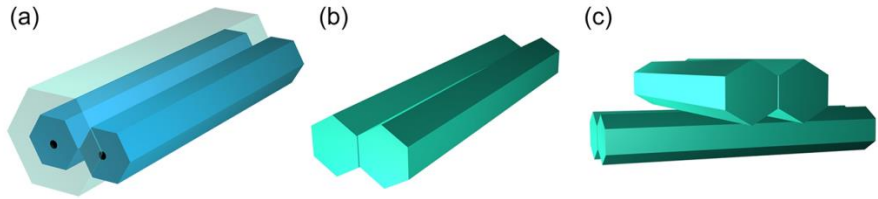

**Fig. 5** Schematic illustration of the origins of the metastable multi-axis orientational ordering of UCNRs. (a) The excluded volume of colloidal rods with hexagonal cross-sections depends on mutual orientations of side faces of the nanorods. (b) The minimized cumulative excluded volume for two hexagonal UCNRs for the two side faces oriented parallel to each other. (c) Two pairs of rods, where side faces are parallel within each pair of UCNRs note that the overlap of excluded volumes between the two pairs of UCNRs in initial orthogonal orientations is very low, which may be the origin of the observed metastable colloidal self-assemblies with multi-axis nematic and smectic-membrane-like ordering.

# Supplementary Information

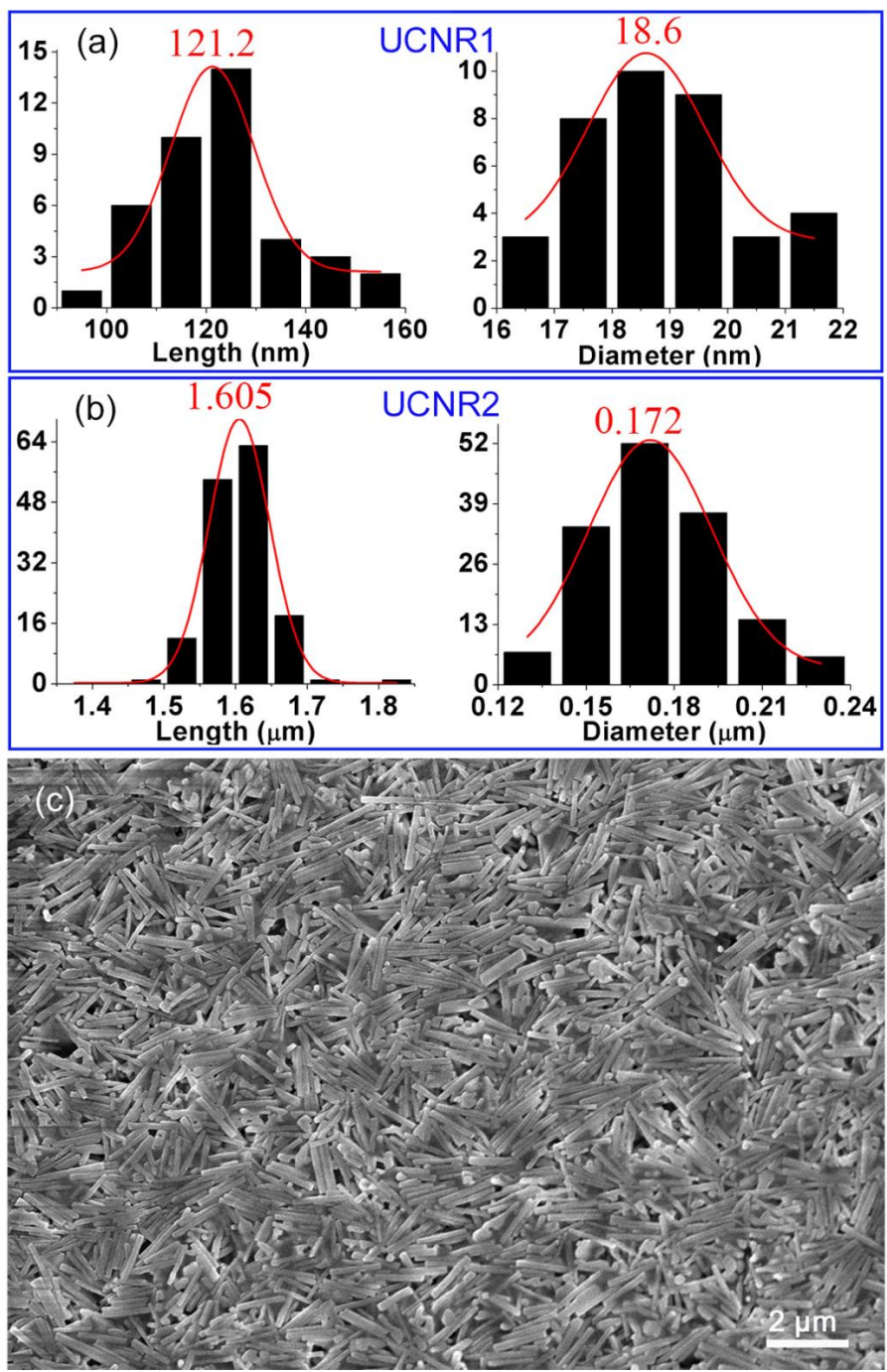

**Fig. S1** Size distributions of (a) UCNR1 and (b) UCNR2 obtained through the analysis of TEM images, such as the ones shown in Fig. 1a, d, and (c) a representative SEM image of the isotropic area of UCNR2. The red lines are the Gaussian fits, using which we characterize the length and diameter of UCNR1s to be 121.2 nm and 18.6 nm, respectively, with an aspect ratio of 6.5. The corresponding length and diameter of the UCNR2s are 1605 nm and 172 nm, with the aspect ratio of 9.3.

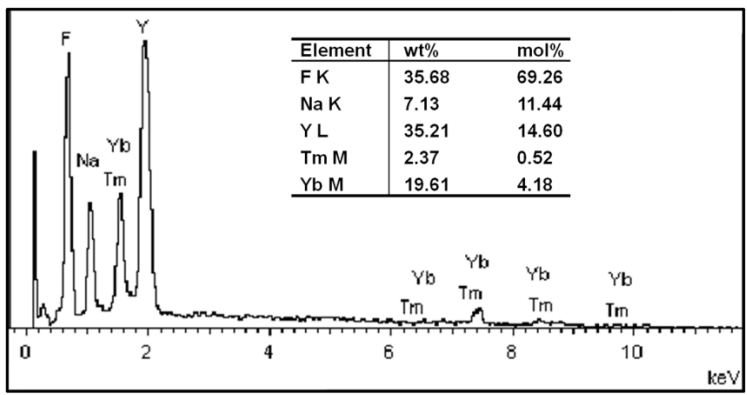

**Fig. S2 EDX microanalysis showing the elemental composition of the UCNR2 nanorods.** Elemental composition and content of Na, Y, Yb, Tm and F are identified on the basis of this analysis. According to the mole percentage, we can evaluate the volume fraction from the obtained experimental results. First, we know that the ratio of valence state of each element is near 1:1 from the measured EDX mol%. The fundamental crystal matrix structure of the UCNR2s is β-NaYF$_4$. Considering that the doped elements Yb$^{3+}$, Tm$^{3+}$ are all cations, and only cations replace the site of cations in the synthesis process, we can obtain the effective UCNR2 composition formula as Na$_{0.66}$Yb$_{0.24}$Tm$_{0.03}$Y$_{0.84}$F$_4$, which is based on the ratio of mol% of each element. Therefore, the effective molecular weight $M^*$ is obtained about 212.46 g/mol.

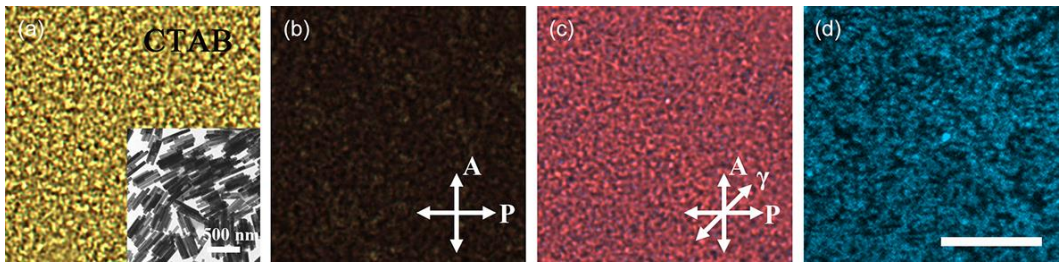

**Fig. S3 Multi-domain LC organization of UCNR1 rods obtained using a micellar surfactant CTAB.** (a)-(d) optical images taken under a conventional optical microscope (a), POM (b), POM with a full-wavelength (530 nm) retardation plate with its slow axis marked by γ (c) and fluorescent microscopy (d), respectively. The UCNR1-CTAB (1.5 mM) binary mixtures exhibit multi-domain orientational ordering, which (unlike in the case of dextran) is spatially inhomogeneous. The scale bar in the optical micrographs, all showing the same sample area, is 30 μm. Inset in (a) provides insights into the details of such ordering revealed by TEM imaging.

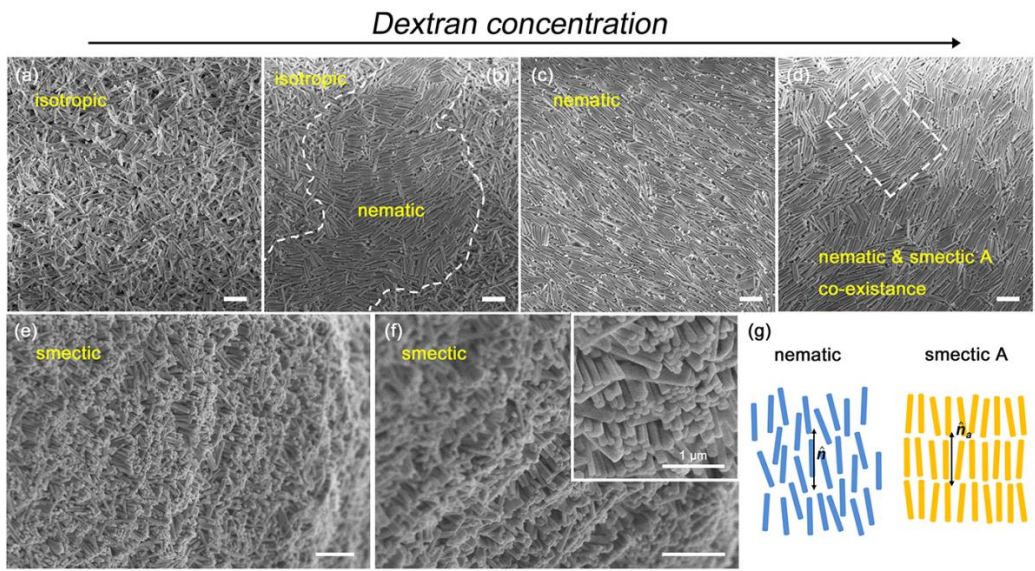

**Fig. S4 SEM images and schematics showing UCNR self-assembled superstructures.** The image sequence reveals the consecutive transitions from isotropic (a) to nematic (b, c) ordering upon the increase of Dextran concentration, as well as the coexistence of nematic and smectic domains (d) that also appears with the increase of dextran concentration. The dashed lines are for guiding the eye and to highlight the coexistence of different types of colloidal organization. (e,f) SEM images showing the cross-sections of the smectic-like domains. (g) schematics of the nematic (left) and smectic A (right) colloidal ordering. The scale bar is 2 μm.

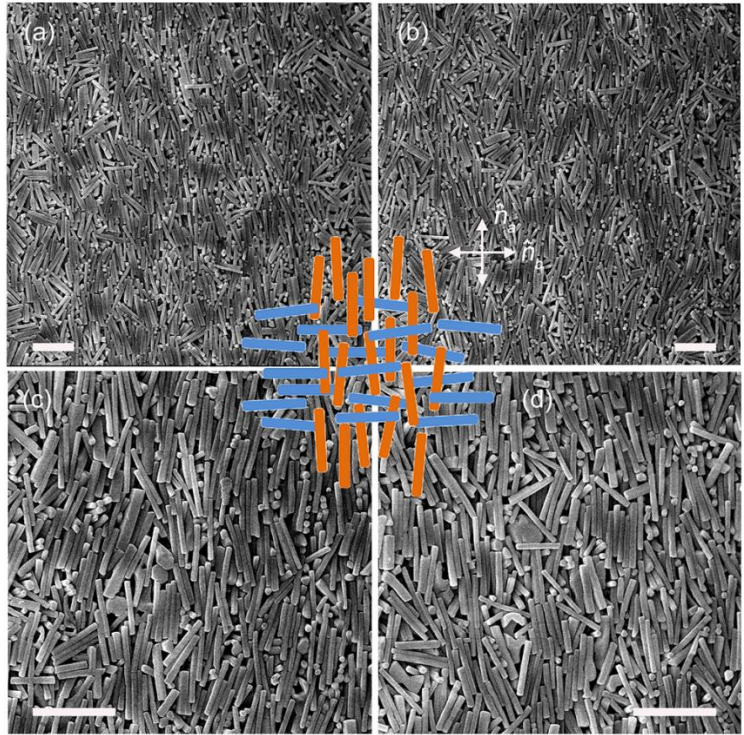

**Fig. S5 SEM images showing the multi-axis nematic ordering of the UCNR2s.** The inset image is the schematic of such multi-axis ordering. The scale bars are all 2 μm.

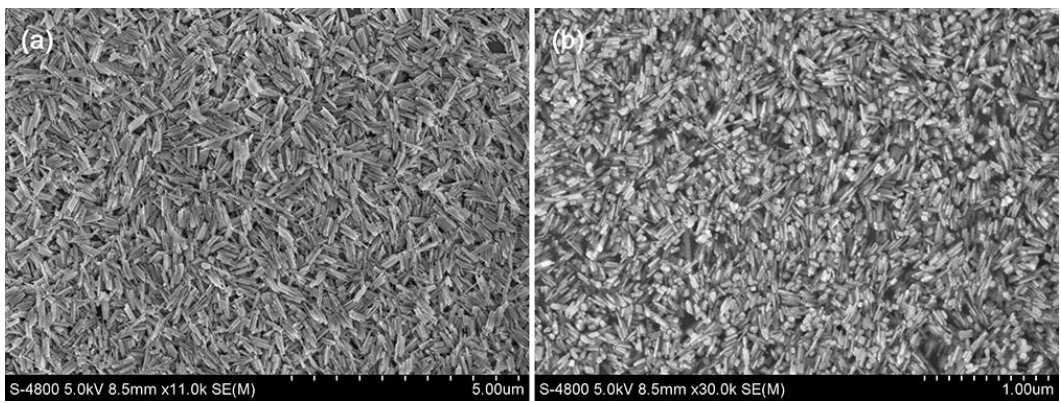

**Fig. S6 Examples of SEM images (a,b) of the colloidal superstructures formed by UCNR1 particles in dispersions with CTAB micelles used as depletants.**